\documentclass[pss]{wiley2sp} 
\usepackage{amsmath}
\usepackage{bm}             

\begin{document}

\title{Time-Dependent Density Functional Theory for Fermionic Superfluids: from Cold Atomic Gases, to Nuclei and Neutron Stars Crust}

\author{%
Aurel Bulgac}

\mail{e-mail
  \textsf{bulgac@uw.edu}}

\institute{%
  Department of Physics, University of Washington, Seattle, WA 98195-1560, USA}


\abstract{\bf%

In cold atoms and in the crust of neutron stars the pairing gap can reach values comparable with the Fermi energy. While in nuclei the neutron gap is 
smaller, it is still of the order of a few percent of the Fermi energy. The pairing mechanism in these systems is due to short range attractive interactions 
between fermions and the size of the Cooper pair is either comparable to the inter-particle separation  or it can be as big as a nucleus, which is still 
relatively small in size. Such a strong pairing gap is the result of the superposition of a very large number of particle-particle configurations, which 
contribute to the formation of the Copper pairs.  These systems have been shown to be the host of a large number of remarkable phenomena, in which 
the large magnitude of the pairing gap plays an essential role: quantum shock waves, quantum turbulence, Anderson-Higgs mode, vortex rings, domain 
walls,  soliton vortices, vortex pinning in neutron star crust, unexpected dynamics of fragmented condensates and role of pairing correlations in 
collisions on heavy-ions, Larkin-Ovchinnikov phase as an example of a Fermi supersolid, role pairing correlations control the dynamics of fissioning 
nuclei, self-bound superfluid fermion droplets of extremely low densities. 
   }

\maketitle   

\section{Introduction.}

There is a large class of fermionic superfluid systems in which the pairing correlations are very strong and their description and especially 
the description of their dynamics and interaction with typically strong external probes require and extension of the Density Functional Theory 
(DFT)~\cite{Dreizler:1990} following the \`a la Kohn and Sham formulation~\cite{Kohn:1965}, which does not involve non-local potentials, 
as in the first extension suggested by 
Oliveira et al.~\cite{Oliveira:1988}. If the pairing gap is large, the 
number of particle-particle configurations defining the anomalous density is much larger than the number $N$ of fermions in the system. In the Kohn-Sham
version of the DFT the energy density functional (EDF) depends on mainly two types of densities, the number density and the kinetic energy density, 
which are expressed through $N$ single particle wave functions (\`a la Hartree-Fock approximation). Since 
in the language of the number density alone one cannot distinguish between normal and superfluid phases, there is an obvious need to 
introduce the anomalous density as well, which defines the order parameter, non-vanishing only in the superfluid phase. However, the description of 
fermionic superfluids becomes even more demanding, since in practice one has to study very often superfluids in interaction with strong time-dependent 
probes, e.g. when one is stirring a fermionic superfluid, when studying non-equilibrium phenomena such as quantum turbulence, and when one may 
even observe the evolution of superfluid into a normal phase.  Naturally, under such circumstances one needs another ``order parameter,'' capable to 
disentangle parts of the system which evolve in time at various rates, 
and in this case the appearance of current densities in the EDF becomes
unavoidable. The physical systems of interests run the gamut from cold atom system to nuclear systems and one has to consider 
multi-component systems for which the structure the EDFbecomes quite complex. In nuclei one has to consider 
at the same time both the (charged) proton and neutron miscible 
superfluids and in neutron star crust in addition also include the electron background as well. In the case 
of cold atoms one is interested lately in miscible 
mixtures of either Fermi-Bose, Fermi-Fermi or Bose-Bose superfluids. In neutron stars various mesons, which 
are bosons, are expected to appear at relatively large densities, close to the core of the star.

\section{Why is there a need for an extension of DFT to time-dependent and superfluid?}

\begin{figure}[t]%
  \includegraphics[width=0.45\textwidth]{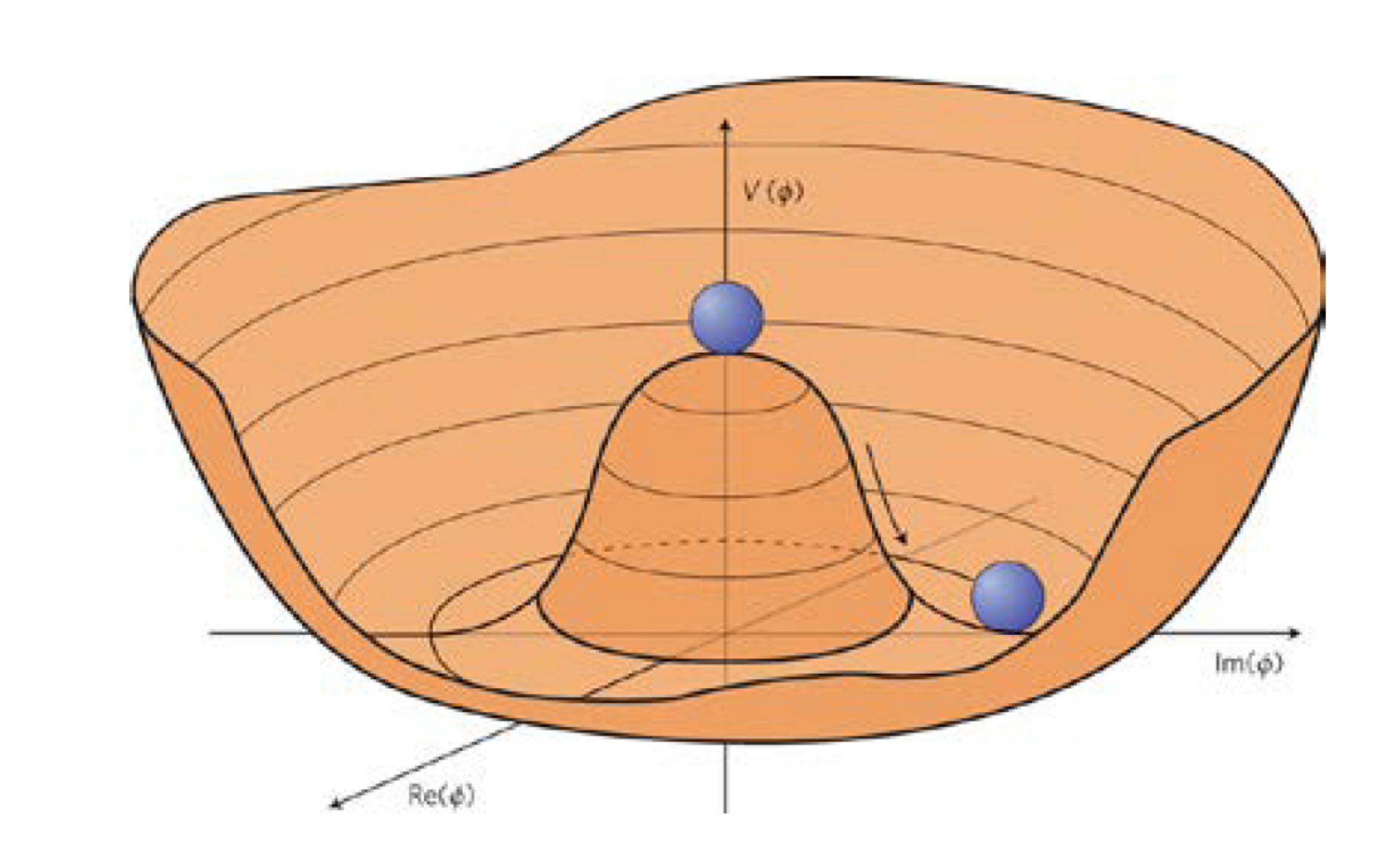}%
  \caption{    \label{fig:mexican_hat}
A rather exotic excitation of a superfluid exist, the Anderson-Bogoliubov-Higgs mode, which corresponds to the amplitude oscillation of 
the order parameter~\cite{Nobel:2013}. Surprisingly the evolution of the magnitude of this mode in time is unlike the motion of a ball rolling.}
 
\end{figure}

Two prevailing theoretical models are used to describe the dynamics of superfluids. The oldest is the Landau two-fluid 
hydrodynamics~\cite{Landau:1941a,Khalatnikov:2000}, which at zero-temperature reduces to the hydrodynamics of a single perfect classical 
fluid~\cite{Lamb:1945,LL6:1966}, namely of the superfluid component alone. Naturally in such a formalism Planck's constant is absent and
the two-fluid hydrodynamics is unable to describe the formation of quantized vortices~\cite{Feynman:1955}, their dynamics, their crossing and 
recombination, which is at the heart of the venerable field of quantum turbulence~\cite{Vinen:2002} conjectured by Feyman in 1955. 
Classical turbulence is due to viscosity, which is absent in superfluids at zero-temperature. On the same note, there is no mechanism 
within the two-fluid hydrodynamics to describe either the
conversion of the superfluid into the normal component, when a superfluid is stirred vigorously.  The quantizations of vortices in  the two-fluid 
hydrodynamics has to be enforced by hand~\cite{Khalatnikov:2000}, in a manner similar to the Bohr 1913  quantization of the hydrogen atom. 
It is thus impossible to describe the evolution of a superfluid at rest and brought into rotation, when quantized vortices are formed, 
and later on when they might cross and reconnect as well.

An extremely attractive alternative approach was developed by 
Gross~\cite{Gross:1961} and Pitaevskii~\cite{Pitaevskii:1961} to describe a weakly interacting Bose 
at zero-temperature, the celebrated Gross-Pitaevskii equation.  This is a non-linear Schr\"odinger equation in which both the density and the 
superfluid order  parameter are described by the same complex field $\psi({\bf r},t)$:
\begin{eqnarray}
i\hbar\frac{\partial \psi({\bf r},t)}{\partial t} &=&- \frac{\hbar^2}{2m}{\bm \nabla}^2 \psi({\bm r},t) \\
&+& g\left |\psi({\bm r},t)\right |^2 \psi({\bm r},t) + V_\text{ext}({\bf r},t ) \psi({\bm r},t), \nonumber 
\end{eqnarray} 
where $g>0$ is the strength of the weak interparticle repulsion and $V_\text{ext}({\bf r},t ) $ is some external potential. While this 
fully quantum mechanical formalism is adequate to describe a large range of properties of a weakly repulsive Bose gas, in particular 
the formation of quantized vortices, it is clearly inadequate to describe the properties of superfluid liquid $^4$He, which is a strongly 
interacting system. Under the influence of a strong time-dependent external field a weakly interacting Bose gas can become normal, at least in
some regions of space, but the Gross-Pitaevskii equation is  unable to disentangle between the normal and the superfluid components.

Zaremba, Nikuni, and Griffin\cite{Zaremba:1999} presented a nice solution to many 
of these issues in the case of weakly interacting bosons. They coupled the quantum Gross-Pitaevskii 
equation describing the condensate, with a classical kinetic equation for the bosons in the normal 
component, which includes collision between non-condensed bosons and also collision 
between the condensed and non-condensed bosons. Landau two fluid hydrodynamics 
emerges when the frequency of the mode satisfies the condition $\omega\tau\ll 1$, 
where $\tau$ is the relaxation time   associated with the collisions between the condensed 
and non-condensed bosons. This new framework is valid only for weakly interacting bosons. 
Later on other more sophisticated approaches have been suggested  for the weakly interacting Bose systems~\cite{Proukakis:2008}.

\begin{figure}[t]%
  \includegraphics[width=0.45\textwidth]{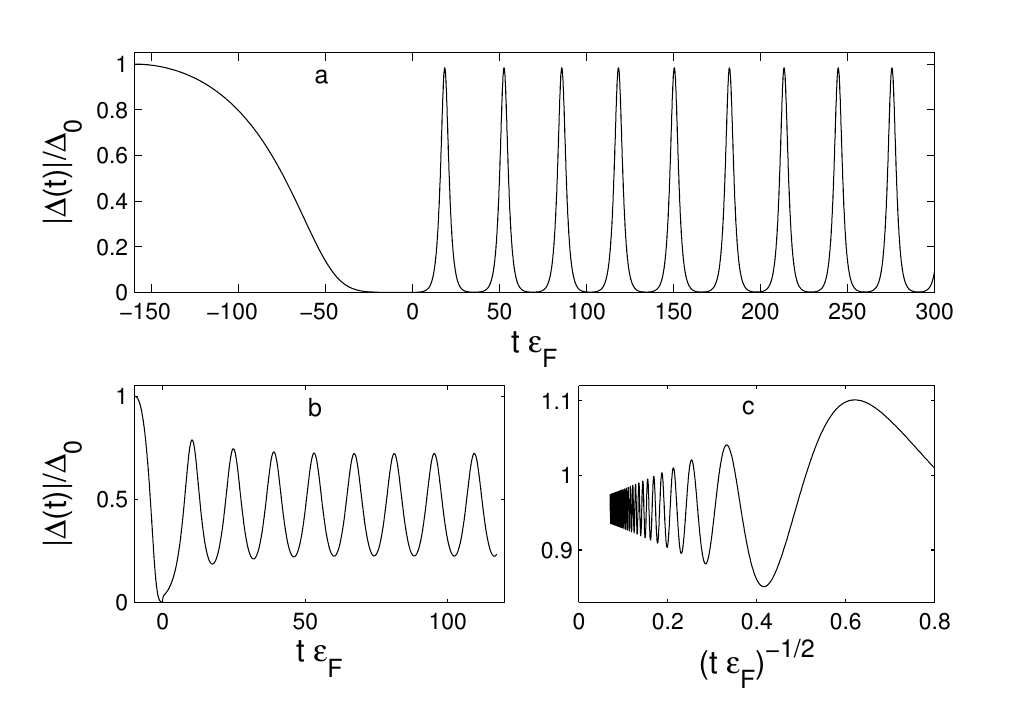}%
  \caption{    \label{fig:AH}
   While exciting the Anderson-Bogoliubov-Higgs mode in an uniform Fermi superfluid 
   the number density remains constant, while the magnitude of the pairing gap and 
the phase change in time, but not in space. 
 Unlike a ball rolling down the hill in Fig. \ref{fig:mexican_hat}, 
the magnitude of the order parameter will oscillate only between zero and the 
equilibrium value, $0< |\Delta(t)|< |\Delta_0|$ ~\cite{Bulgac:2009b}. There is no such 
collective  mode in either Landau two-fluid hydrodynamics or Gross-Pitaevskii equation.  
 The  unitary Fermi gas is left to evolve freely in time with the TDDFT extended to superfluid system, after 
 the magnitude pairing gap is brought very slowly out of equilibrium to either a relatively small value at $t=0$, panels $a$ and $b$, 
 or only slightly changed from its equilibrium value, panel $c$. (Here $\varepsilon_F$ is the Fermi energy and $\hbar=1$.)
 In cold Fermi gases one can excite these modes by means of the 
 Feshbach resonance~\cite{Zwerger:2011}, by preparing the system with 
 a value of the scattering length close to the BCS limit and suddenly changing it to unitarity at $t=0$ . 
 In a uniform Fermi gas by controlling the magnitude of the scattering length 
 one can control the magnitude of the pairing gap while maintaining the number density constant.  
 }
\end{figure}

Apart from these difficulties discussed above, none of these models of superfluids allow for the existence of the 
Anderson-Bogoliubov-Higgs mode. 
This mode was noticed by P. W. Anderson a long time ago~\cite{Anderson:1958,Anderson:2015,Littlewood:1982}. The potential energy of a system with a 
complex order parameter has a shape similar to a Mexican hat, see Fig. \ref{fig:mexican_hat} taken from Ref.~\cite{Nobel:2013}). 
Typically this potential depends only on the 
magnitude of the complex field but not on its phase, e.g. $V(\phi)=a|\phi |^2+b|\phi |^4$, where $a<0$ and $b>0$ and a minimum value at 
$|\phi_0|^2=-b/2a$.  The mode $|\phi_0|$ is 
known in high-energy physics as the Higgs boson and it became an essential element of the 
Standard Model. The existence of the Higgs boson leads to masses of quarks, gluons, electrons, Z$^0$, 
and W$^\pm$ bosons, and other elementary particles and its existence has been 
determined experimentally~\cite{Nobel:2013}. 

If in a homogeneous unitary Fermi Gas (see next section for its characterization) 
one would bring very slowly the pairing gap out of the equilibrium position $\Delta_0$  
in the ground state one would be able to observe at least two kinds of rather unexpected excitation modes. 
The first kind of excitation correspond to small amplitude oscillations around the equilibrium, 
with an unexpected slow algebraic damping~\cite{Volkov:1974,Bulgac:2009b}, when the magnitude of 
the pairing gap behaves as a function of time as 
\begin{equation} 
|\Delta(t)=|\Delta_\infty|+ \frac{A}{\sqrt{2|\Delta_\infty|t}}\sin(2|\Delta_\infty|t+\theta), \label{eq:volkov}
\end{equation}
where $|\Delta_\infty|<|\Delta_0$. The asymptotic state corresponds to a partially fermionic paired state plus quasi-particle excitations.
 If instead the pairing gap is slowly brought to a very small value and after that the system is left 
to evolve freely, the magnitude of the pairing gap oscillates with a maximum value smaller than the equilibrium value, 
$0<|\Delta(t)|<|\Delta_0|$, see Fig. \ref{fig:AH}.  
The same results illustrated in Fig. \ref{fig:AH} can be 
obtained by preparing initially the system with an interaction strength 
corresponding to an equilibrium pairing gap smaller than $|\Delta_0|$ and suddenly changing the coupling strengths to a value 
corresponding to an equilibrium value of the pairing gap $|\Delta_0|$. 

\section{Energy density functional for the unitary Fermi gas.}
We will illustrate the extension of the Kohn-Sham local density approximation (LDA) of the DFT to superfluid fermionic systems at first 
with the case of  an unitary Fermi gas, which is both methodologically clear and of great practical value. 
In a subsection section we will later briefly discuss the structure of the EDF for nuclei and neutron stars. 

As in the case of normal electron systems, one of the most frustrating aspects of DFT is the construction of the EDF, 
which we know it exits according to the Hohenberg-Kohn theorem~\cite{Hohenberg:1964}, 
but for which however we have no well defined procedure to 
construct with enough high accuracy. The unitary Fermi gas~\cite{Baker:1999,Zwerger:2011}, which is a system of spin-up and down fermions,
interacting with a zero-range potential, characterized by an infinite scattering length and a zero range effective become an object of extremely intensive
study both experimentally and theoretically in the last two decades. George Bertsch noticed~\cite{Bertsch:1999} 
that the neutron matter in the crust of neutron stars is very close
to such an idealized system. 

Neutron interaction in the $s$-wave is characterized by a very large scattering length $a$ and a relatively small effective 
range $r_0$, with an low-energy $s$-wave scattering amplitude
\begin{equation} \label{eq:f-s}
f=\frac{1}{-ik-\tfrac{1}{a}+\tfrac{1}{2}r_0k^2+\ldots},
\end{equation}
where $k$ is the relative wave vector of two scattering fermions. The wave function outside the potential is
\begin{equation}
\label{eq:wf}
\psi({\bm r}) = \exp( {\bm r}\cdot{\bm k}) + \frac{f}{r}\exp(ikr)\approx 1 -\frac{a}{r}+{\cal O}(kr).
\end{equation}
The total scattering cross section of two low-energy fermions $\sigma =4\pi |f|^2 \rightarrow \tfrac{4\pi}{k^2}$ reaches the maximum 
possible value allowed by unitarity if $r_0 \rightarrow 0$ and $|a|\rightarrow \infty$. 
If the scattering length $a$  is positive then a bound state exist with the wave function 
\begin{equation}
\psi({\bm r}) =\frac{1}{\sqrt{2a}r}\exp\left (-\frac{r}{a}\right )+{\cal O}\left (\frac{r_0}{a}\right ). \label{eq:wfbs}
\end{equation}
If in a many-fermion system, in which the average interparticle separation is $\propto n^{-1/3}$,  the conditions, 
\begin{equation}
\label{eq:ufg}
r_0 \ll n^{-1/3} \ll |a|
\end{equation}
are satisfied such a system is called a unitary Fermi gas (UFG)~\cite{Zwerger:2011}. 

In 1999  the dilute neutron matter (for which the condition \eqref{eq:ufg} is weakly satisfied) 
was he closest physical system to an UFG one could envision, and the calculation of the value of the dimensionless Bertsch parameter $\xi$  
became a theoretical challenge. If $\xi<0$ the system would collapse into a high density liquid or solid with an average interparticle separation likely 
of the order of the range of the interaction and with properties determined by the particular features of the interaction between fermions. 
However, if $0<\xi\le 1$ the ground state would be that of a gas, and a very unusual gas at that, a superfluid 
with a pairing gap of the order of the Fermi energy, the largest pairing gap in any physical system, and universal
properties largely independent of the details of the interaction.
One can easily establish using dimensional arguments alone that the
ground state energy of a uniform UFG should be given by a function depending only on the volume $V$, particle number $N$, $\hbar$, 
and the fermion mass $M$, in the unitary limit when $r_0\rightarrow 0$ and $|a|\rightarrow \infty$:
\begin{equation}
\label{eq:e_ufg}
E_\text{gs}(N,V,\hbar, m, a, r_0)\rightarrow \xi N\frac{3}{5} \varepsilon_F, 
\end{equation} 
where $\varepsilon_F=\tfrac{\hbar^2k_F^2}{2m}$ is the Fermi energy of a uniform non-interacting Fermi gas with the Fermi wave-vector 
$k_F = \left ( 3\pi ^2\tfrac{N}{V}\right )^{1/3}$, and $\xi$ is the dimensionless Bertsch parameter.  $\xi\le 1$,  
as the scattering length $a$ can become infinite 
only in the case of  attractive interactions. Apart from the universal value of the Bertsch parameter $\xi$, the EDF for the UFG 
should have the same structure as for a non-interacting Fermi gas and the only parameter specifying the nature of the fermions is their mass. 
Both theoretical Quantum Monte Carlo (QMC) $\xi=0.372(5)$~\cite{Carlson:2011} and experimental $\xi=0.376(5)$~\cite{Ku:2011} values of the 
Bertsch parameter are now in very good agreement with each other. 

A fermionic superfluid under the influence of a time-dependent external field might become normal in some spatial regions  but not in others, 
and the number density alone cannot discriminate between different phases.
The case of the UFG is particularly attractive from the point of view of a DFT aficionado, as only rather general requirements 
suffice to narrow down the structure of the EDF. Dimensional arguments, rotation, translational, and parity symmetries, gauge symmetry related 
to the transformations of the complex order parameter, Galilean covariance, and renormalization and regularization of the EDF combined with the so 
called adiabatic local density approximation (ALDA)~\cite{Gross:2012} extended to the case of superfluid systems restrict the form of the (unregulated) 
EDF for UFG to a rather simple form~\cite{Bulgac:2007a,Bulgac:2011,Bulgac:2013b}, namely:
\begin{eqnarray}
&&{\cal E}({\bm r},t)=\frac{\hbar^2}{2m}  \left [ \alpha \tau({\bm r},t) + \beta \frac{3}{5} (3\pi^2)^{3/2}  n^{5/3}({\bm r}, t) \right . \label{eq:eos_ufg}\\
&&+ \left . \gamma \frac{|\nu({\bm r},t)|^2}{n^{1/3}({\bm r},t)} +(1-\alpha)\frac{{\bm j}^2({\bm r},t)}{n({\bm r},t)}  \right ] +V_\text{ext}({\bm r},t)n({\bm r},t),\nonumber
\end{eqnarray} 
where $\alpha$, $\beta$, and $\gamma$ are dimensionless constants. 
$n({\bm r}, t)$, $\tau({\bm r},t)$, $\nu({\bm r},t)$, and ${\bm j}({\bm r},t)$ are the unregulated number, kinetic, anomalous densities
and current densities of a fully unpolarized UFG 
(equal number of spin-up and spin-down particles) and  expressed through the 
Bogoliubov quasiparticle amplitudes $u_n({\bf r},t)$ and $v_n({\bf r},t)$~\cite{Gennes:1966}.   $V_\text{ext}({\bm r}t)$ 
is an arbitrary external field with which one might probe or excite the system. We refer to this form of DFT 
as the (Time-Dependent) Superfluid Local Density Approximation ((TD)SLDA), which is a natural extension of the LDA for normal 
systems of Kohn and Sham formulation of the DFT~\cite{Kohn:1965} to superfluid systems.
The emerging TDSLDA equations have the expected form, identical to the Bogoliubov-de Gennes equations, 
\begin{equation}
\label{eq:eq_ufg}
i\hbar \frac{\partial}{\partial t} 
                    \left ( \begin{array}{c}  u_n \\ v_n \end{array} \right ) 
                  =\left ( \begin{array}{cc}  h-\mu  &\Delta\\
                                                         \Delta^*& -h^*+\mu  \end{array} \right )                  
                        \left ( \begin{array}{c}  u_n \\ v_n \end{array} \right ), 
\end{equation}
where $h=\tfrac{\partial {\cal E}}{\partial n}$, $\Delta=\tfrac{\partial {\cal E}}{\partial \nu^*}$, and $\mu$ are the single-particle Hamiltonian, 
the pairing field, and the chemical potential. 
Both kinetic energy and anomalous densities 
diverge as a function of the ultraviolet cutoff
in a very similar manner in the case of a zero-range interaction~\cite{Bulgac:2011}.
In particular the anomalous density matrix has the behavior
\begin{equation}
\nu({\bm r}_1,{\bm r}_2,t) \propto \frac{1}{ |{\bm r}_1-{\bm r}_2| }, 
\quad \text{if} \quad  
|{\bm r}_1-{\bm r}_2| \rightarrow  0.
\end{equation}
This divergence is the same as the divergence of the scattering wave in Eqs. \eqref{eq:wf} or 
of the bound state wave function \eqref{eq:wfbs} when $r\rightarrow r_0 \rightarrow 0$. This divergence is ``real'' and 
not a deficiency of the theoretical formalism. 
The divergence of the anomalous density matrix reflects nothing else but the increase of the wave function 
of the Cooper pair when the separation between the fermions approaches the radius of the interaction. 
One can relate the anomalous density matrix $\nu({\bm r}_1,{\bm r}_2,t)$ with the Cooper pair wave function.  

The mathematical difficulty in extending DFT in \'a la Kohn and Sham manner to superfluid system  
arises when one attempts to reach the limit $n^{1/3}r_0\rightarrow 0$, when the average interparticle separation is much larger 
than the radius of the interaction  and avoid 
the appearance of infinities in the calculations of various densities and of the pairing gap. 
In meanfield approximation the pairing field $\Delta({\bm r}_1, {\bm r}_2,t)= -V({\bm r}_1-{\bm r}_2)\nu({\bm r}_1,{\bm r}_2,t) $, 
where $V({\bm r}_1-{\bm r}_2)$  is the fermion-fermion interaction responsible for the pairing correlations, is formally non-local.
In the limit $n^{1/3}r_0\rightarrow 0$ densities should be calculated with a cutoff~\cite{Bulgac:2002,Bulgac:2011}:
\begin{eqnarray}
n({\bm r}, t)&=&  2 \sum_{0<E_n<E_\text{cut}} |v_n({\bf r},t)| ^2 ,\\
\tau({\bm r},t)&=&2\sum_{0<E_n<E_\text{cut}} |{\bm \nabla} v_n({\bf r},t)| ^2 ,\\
\nu({\bm r},t)&=&     \sum_{0<E_n<E_\text{cut}} v^*_n({\bf r},t)u_n({\bm r},t)                     ,\\
{\bm j}({\bm r},t)&=&  2\; {\text {Im}}\hbar\sum_{0<E_n<E_\text{cut}} v^*_n({\bf r},t) {\bm \nabla} v_n({\bm r},t),
\end{eqnarray}
in which the pre-factor 2 is for the spin multiplicity. 
In time-dependent simulations one usually starts with the system prepared in its ground state and the initial quasi-particle amplitudes  
$u_n({\bf r},t)=u_n({\bf r})\exp\left(-i\tfrac{E_nt}{\hbar}\right)$ and $v_n({\bf r},t)=v_n({\bf r})\exp\left(-i\tfrac{E_nt}{\hbar}\right)$. 
$E_n$ are the eigenvalues of the initial SLDA stationary equations and $E_\text{cut}$ is an ultraviolet cut-off energy, which when 
chosen large enough does not affect the values of any physical observables~\cite{Bulgac:2002}.
In the ground state the single-particle Hamiltonian of an unpolarized UFG
has the structure
\begin{equation}
h = -\alpha \frac{\hbar^2}{2m}\Delta + U({\bf r}).
\end{equation}
One needs to introduce the momentum dependent wave-vectors and the renormalized coupling constant
\begin{eqnarray}
&&\alpha \frac{\hbar^2k_0({\bm r})}{2m}+U({\bm r})-\mu=0,\\
&&\alpha \frac{\hbar^2k_c({\bm r})}{2m}+U({\bm r})-\mu=E_\text{cut},\\
&&\frac{1}{g_\text{eff}({\bm r}) }=  \frac{ n^{1/3}({\bm r}) }{\gamma} \label{eq:g}\\
&& - \frac{ mk_c({\bm r} )}{ 2\pi^2\hbar^2 \alpha}
\left [ 1 - \frac{ k_0({\bm r}) }{ 2k_c({\bm r}) } 
\ln \frac{ k_c({\bm r})+k_0({\bm r}) }{ k_c({\bm r})-k_0({\bm r}) } \nonumber 
\right ],
\end{eqnarray}
in roder to derive the renormalized form of the pairing gap
\begin{equation}
\Delta({\bm r})= -g_\text{eff}({\bf r})\nu({\bm r}).
\end{equation}
One can then show that the combination
\begin{equation}
\alpha\frac{\hbar^2}{2m}\tau({\bm r}) -\Delta({\bm r})\nu^*({\bm r})
\end{equation}
does not diverge when $E_\text{cut}\rightarrow \infty$. While the wave vector $k_0({\bm r})$ can become imaginary in the classically
forbidden regions of space, the effective coupling consnat $g_\text{eff}({\bm r})$ remains real. 
If $k_c({\bm r})$ becomes imaginary in any spatial region the recipe is that the last term on the rhs side of 
Eq. \eqref{eq:g} should be dropped~\cite{Bulgac:2002,Bulgac:2011}.

The EDF for the UFG Eq. \eqref{eq:eos_ufg} depends on three dimensionless parameters $\alpha, \beta$ and $\gamma$, 
for both superfluid and normal phases, which can 
be extracted from values of the Bertsch parameter $\xi$, the pairing gap, and the momentum dependence of the quasi-particle excitations obtained 
in the QMC for the uniform UFG~\cite{CCPS:2003,Carlson:2005,Carlson:2011,Chang:2004,ABCG:2004,Lobo:2006},
which agree well with extracted experimental values~\cite{Carlson:2008,Schirotzek:2008,Ku:2011}.

\begin{figure}%
  \includegraphics[width=0.45\textwidth]{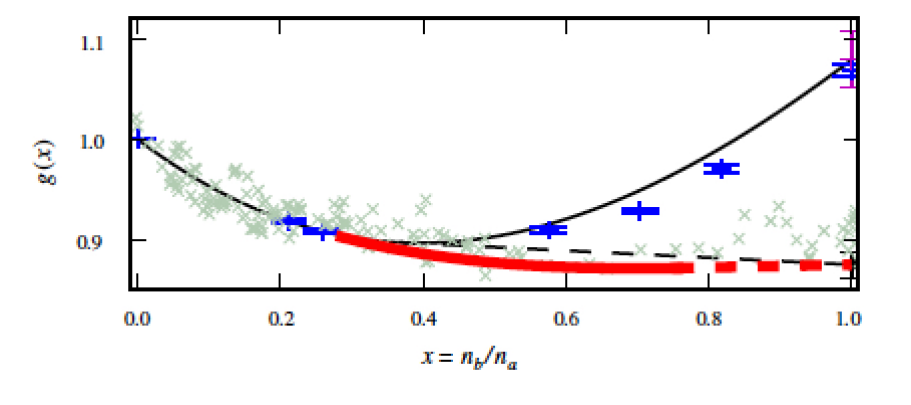}%
  \caption{    \label{fig:uUFG}
The function $g(x)$ can be extracted from QMC and 
other theoretical calculations~\cite{Lobo:2006,Pilati:2008,Combescot:2007}, 
blue points with error bars for the normal state of UFG and the black point 
with error bars at $x=1$ for the unpolarized superfluid UFG) and also directly 
from experiment~\cite{Shin:2008} (green crosses).  
The black dashed line represents the mixed phase discussed in 
Refs.~\cite{Lobo:2006,Pilati:2008,Combescot:2007}. The red line is for the Larkin-Ovchinnikov (LO)
phase: the solid line is for the pure state and the dashed line is for the mixed state 
with the superfluid unpolarized state\cite{Bulgac:2008}. At the left end of the red solid 
line there is a second order phase transition between the normal and the LO phase, 
while at the right end of the red solid line there is first order transition between the 
pure LO phase and the mixed phase. Unlike in the case of weakly BCS superfluids, 
the LO phase exits at unitarity in a very wide window of polarization $x=n_b/n_a$. }
\end{figure}

In the case of polarized UFG one needs two number densities $n_{a,b}({\bm r}, t)$ and from dimensional 
arguments alone it follows  that the energy density of a uniform polarized UFG is:
\begin{equation}
{\cal E}(n_a,n_b)= \frac{3}{5}\frac{\hbar^2}{2m} (6\pi^2)^{2/3} \left [n_ag\left(\frac{n_b}{n_a} \right )\right]^{5/3}
\end{equation}
with $g(1)=(2\xi)^{3/5}$, see Fig. \ref{fig:uUFG}.  

\subsection{Validation of (TD)SLDA.} At this point one is in the position to validate 
the accuracy of the suggested form of the EDF for the UFG \eqref{eq:eos_ufg} by placing various
number of fermions with spin-up and spin-down in external fields. 
A UFG in a harmonic trap is particularly interesting, since from theory we know that its properties are determined by the only energy scale in the system 
$\hbar \omega$~\cite{Thomas:2005,Werner:2006,Son:2007}, their properties can be calculated numerically with controlled accuracy within QMC 
and unlike the the homogeneous system density gradients are important.
The EDF for the UFG has so far been 
constructed for uniform systems and it is not obvious that such an EDF 
would perform well for inhomogeneous systems, where density gradients are significant.
Fortunately there exist a sufficiently large number of QMC calculations of  
unpolarized and polarized finite systems of fermions at unitarity in a external harmonic trap, both in the superfluid and normal phases. 
The  EDF for the polarized UFG has a little more complex structure, as in this case there are two independent number densities, for spin-up 
and spin-down fermions. Similar arguments as those used for the derivation of the EDF for an unpolarized UFG can be used for its 
derivation~\cite{Bulgac:2007,Bulgac:2008,Bulgac:2008a,Bulgac:2011}. In case of inhomogeneous systems a new gradient 
term might need to be included in Eq. \eqref{eq:eos_ufg}, namely one of the form 
$\propto\left  |{\bm \nabla} \sqrt{n({\bf r},t)}\right |^2$, but comparison with QMC 
calculations of inhomogeneous systems indicate that it can be neglected~\cite{Bulgac:2007a}. Polarized fermions at unitarity can 
be in either superfluid or normal phase, depending on the degree of polarization 
$N_\downarrow/N_\uparrow$~\cite{Bulgac:2007,Bulgac:2008,Bulgac:2008a,Bulgac:2011} and the same EDF can be used to describe 
with surprisingly high accuracy all such systems. When comparing the results of the QMC calculations performed for trapped unitary 
fermions~\cite{Blume:2007,Stecher:2008}
with the predictions of the SLDA one finds almost perfect agreement~\cite{Bulgac:2007a,Bulgac:2011}. The differences between the 
QMC calculations and the SLDA predictions are almost always within the statistical errors of the 
QMC calculations, within at most 1-3\%. There are two exceptions. 
The QMC and SLDA calculations for the $(N_\uparrow,N_\downarrow)=(15,14)$ have an error of 9.5\%, attributed to 
the inaccuracies in the QMC results for this particular system. The largest disagreement is for the two-fermion system 
$(N_\uparrow,N_\downarrow)=(1,1)$, about 15\%. The rest of the differences between the QMC results and the SLDA predictions are at the 
level of $1\dots 3$ \%, which is also the level of accuracy of the QMC calculations. Surprisingly, the odd-even staggering, namely the energy 
differences between the ground state energies of even $N_\uparrow=N_\downarrow$   and odd systems $N_\uparrow-N_\downarrow=1$ 
are within statistical errors as well.  One should keep in mind the ground state energies of harmonically unpolarized trapped systems of unitary 
fermions scale as $\sqrt{\xi}E_{NI}\propto\hbar\omega N^{4/3},$  where $E_{NI}$ is the energy of $N$ non-interacting fermions 
in a harmonic trap. 
The odd-even energy differences depend relatively weakly on the total particle number~\cite{Blume:2007,Stecher:2008} and 
the relatively small odd-even energies differences are reproduced within SLDA with surprising accuracy as well.

There are only a few exact solutions of the time-dependent Schr\"odinger equation for interacting  many-fermion  systems. 
The linear response theory predicts damped harmonic 
oscillations with a frequency $2|\Delta_0|$~\cite{Anderson:1958,Volkov:1974}, while 
Eq. \eqref{eq:volkov} emerges when nonlinearities are taken into account.
Equally unexpected is the behavior of the pairing gap when the initial disturbance is
large, panels $a$ and $b$ in Fig. \ref{fig:AH}, when one would naively expect that 
the pairing gap would oscillate somehow around the equilibrium value $|\Delta_0|$.  
While these modes have been put in evidence in a meanfield-like framework and 
one might suspect that they are artifacts of possible approximations, 
the same behavior was demonstrated to appear in an exactly soluble time-dependent 
realistic many-body model of superconductivity, the Richardson 
or Gaudin model~\cite{Yuzbashyan:2006}. We will illustrate below however the power of TDSLDA 
in confronting actual non-equilibrium phenomena in real experiments.

\subsection{Nuclear systems.} 
Nuclear systems are significantly more complex that the UFG.  While we know for decades quite 
a bit about the nuclear EDF (NEDF),
the exact from and accuracy of mostly empirical NEDF is still insufficient for many applications, like 
predicting the origin of elements in Universe. In case of nuclear systems there two type of fermions, 
protons and neutrons, and the spin-orbit interaction is very strong, and has the opposite sig when 
compared to atomic systems. Pairing correlations are relatively strong as well, though not as strong 
as in the case of UFG. However, the dilute neutron matter, found in the crust of neutron stars, 
has quite a lot of similarities with the UFG~\cite{Bertsch:1999,Baker:1999}.  QMC calculations 
for nuclear systems a  significantly more complex than for electronic or cold atom systems, and so far
only the pure neutron matter equation of state as function of density is know with reasonable accuracy 
from {\ it ab initio} calculations. The interaction between nucleons (neutrons and protons) is very 
complex and in nuclei not only two-body interactions are important, but also three-body (and even four-body) 
interactions play a crucial role. As a result the form of NEDF is typically obtained with significant 
phenomenological input~\cite{Bender:2003,Shi:2018}. The accuracy of phenomenological NEDFs in predicting the binding energies 
of about  2300 known nuclei is nowadays at the sub-percent level, but still not accurate enough for many applications.

In the resulting evolution equations
we suppressed for the sake of simplicity the space and time coordinates 
$({\bm r}, t)$. The ensuing equations
represent an infinite set of coupled nonlinear time-dependent 3D
partial differential equations for the quasi-particle wave functions, 
 \begin{eqnarray} \label{eq:tdslda}
i\hbar
\left ( \begin{array}{c}
\dot{u}_{k\uparrow}  \\
\dot{u}_{k\downarrow} \\
\dot{v}_{k\uparrow} \\
\dot{v}_{k\downarrow}
\end{array}
\right )
=
\left ( 
\begin{array}{cccc}
h_{\uparrow \uparrow}  & h_{\uparrow \downarrow} & 0 & \Delta \\
h_{\downarrow \uparrow} & h_{\downarrow \downarrow} & -\Delta & 0 \\
0 & -\Delta^* &  -h^*_{\uparrow \uparrow}  & -h^*_{\uparrow \downarrow} \\
\Delta^* & 0 & -h^*_{\downarrow \uparrow} & -h^*_{\downarrow \downarrow} 
\end{array}
\right ) 
\left (
\begin{array}{c}
u_{k\uparrow} \\
u_{k\downarrow} \\
v_{k\uparrow} \\
v_{k\downarrow}
\end{array}
\right).
\end{eqnarray}
Here both the local mean field $h_{\sigma\sigma'}$ and pairing field
$\Delta$ depend on the various single-particle densities. 
The index $k$ labels each quasi-particle
wave function, and is both discrete and
continuous. 
This index must also run over isospin, so that there are
similar sets of equations for both protons and neutrons, which naturally are coupled.  We have
explicitly included the spin indices $(\sigma,\sigma') \in \{\uparrow, \downarrow\}$,
allowing for mixing between the spin-up and spin-down
states by the spin-orbit interaction, thus capturing the effects
of proton-proton and neutron-neutron pairing.

\subsection{Numerical implementation.} The TDSLDA equations 
are discretized in space and time. The system of interest is placed on a spatial lattice
with $N_xN_yN_z$ lattice points, for a chosen lattice constant $l$, which determines the 
momentum cutoff $\hbar \pi/l$~\cite{Bulgac:2013}. When these simulation box parameters
are chosen appropriately one can ensure that with further discretization the corrections are 
(exponentially) small~\cite{Bulgac:2013}. The equations are propagated in time using a 6th order multi-step
predictor-corrector-modifier, which typically requires an application of the 
quasi-particle Hamiltonian only twice per time-step. Notice that in the case of the 
popular Runge-Kutta 4th order method one would need to apply the Hamiltonian four times per time step.
Unitarity and accuracy of various conserved quantities 
during the time evolution are satisfied with high accuracy for up to millions of time steps.  
The number of coupled complex 3D time-dependent nonlinear partial 
differential equations which has to be evolved in time is up to  ${\cal O}(10^6)$, 
see supplemental material in Refs.~\cite{Bulgac:2011,Bulgac:2014a,Bulgac:2018a}. 
The numerical solution of these equations requires the use of 
the leading edge supercomputers and it ranks among some of the largest direct numerical simulations
attempted so far.

\section{SLDA for cold Atomic gases, nuclei, and neutron star crust.}

This section will briefly describe qualitative aspects of static and dynamic aspects of fermionic superfluids in 
strongly interacting cold atomic gases, nuclei and neutron star crust. 

\subsection{Quantized vortex in cold superfluid Fermi gases and dilute neutron matter.}

One of the first applications of the SLDA for UFG was to determine the structure of a quantized vortex~\cite{Bulgac:2003,Yu:2003a}. It was
shown that the actual
number density profile of a quantized vortex is somewhere between that in a BCS superfluid and in a  BEC superfluid. 
While in a BCS superfluid the density in the core is practically the same as the density far away from the vortex core and only the order 
parameter vanishes at the vortex core,  in a BEC quantized vortex both the order parameter and the number density vanish at the core, in  
the case of dilute superfluids. In the case of the UFG while the order parameter vanishes at the core, 
the vortex core is only partially filled with fermions, with a density only about half the value of the asymptotic value. 
This density depletion of the vortex core was used to visualize in experiments the Abrikosov 
vortex lattice formed in a  rotating UFG~\cite{Zwierlein:2005}, which was the deciding experimental 
argument in demonstrating that UFG is indeed a superfluid. 

\subsection{The Larkin-Ovchinnikov phase.} 
Once the EDF of the UFG was established and validated, it was used to in order to establish if a UFG can sustain an inhomogeneous 
state of the order parameter~\cite{Larkin:1964,FF:1964}. In Ref.~\cite{Bulgac:2008} the selfconsistent  
SLDA equations for a system in which the order parameter 
was allowed to oscillate in the $z$-direction, while in the $x$ and $y$ directions the properties of the system remains homogeneous, were solved 
Unlike in the case of a weakly coupled BCS superfluid, where the LO phase exist only for a very narrow window of spin polarization, the UFG
was shown to sustain such a phase in a surprisingly wide range, see Fig.~\ref{fig:uUFG}. Only the amplitude of the oscillations of the number density of 
the minority component are significant. This LO phase has not been observed yet  in cold fermi atom systems. 

\begin{figure}%
  \includegraphics[width=0.45\textwidth]{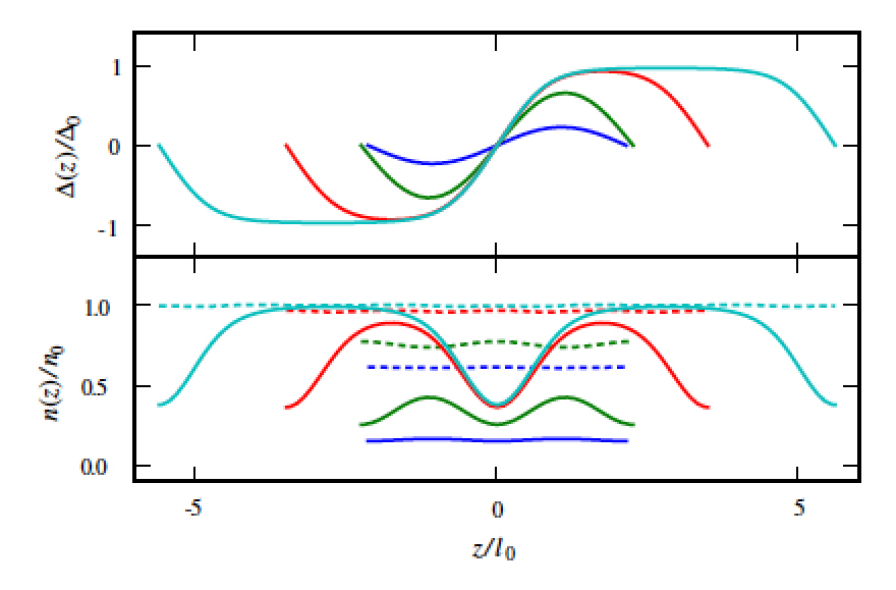}%
  \caption{    \label{fig:LO}
The spatial profile of the pairing gap $\Delta (z)$ and of the number densities $n_{a,b}(z)$ of the majority (dotted) and minority 
species (solid) in the region where a pure LO phase exist, see solid red line in Fig.~\ref{fig:uUFG}. 
For polarization close to the left end of the solid red curve in Fig.~\ref{fig:uUFG}, the spatial 
shape of the order parameter is very similar to a sine-function and the 
amplitude of the oscillation of the order parameter is small (blue curves). Close to the right end of the solid red curve in 
Fig.~\ref{fig:uUFG} the spatial shape of the order parameter starts resembling a domain wall of finite width.
For each polarization the optimal period of the LO phase determined.}
\end{figure}

\begin{figure}[h]

\includegraphics[width=0.45\textwidth]{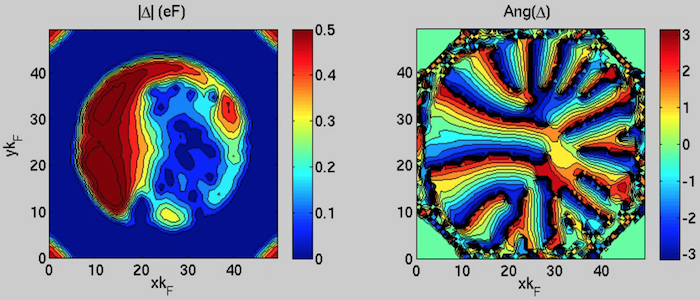}

\includegraphics[width=0.45\textwidth]{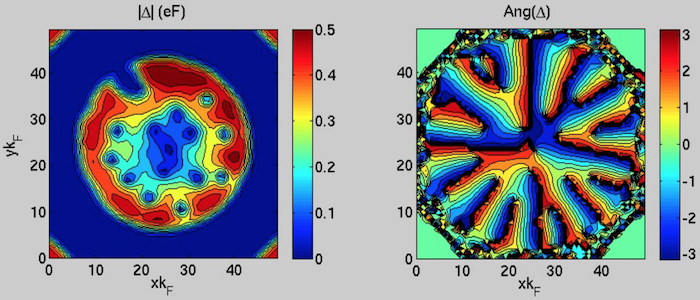}

\includegraphics[width=0.45\textwidth]{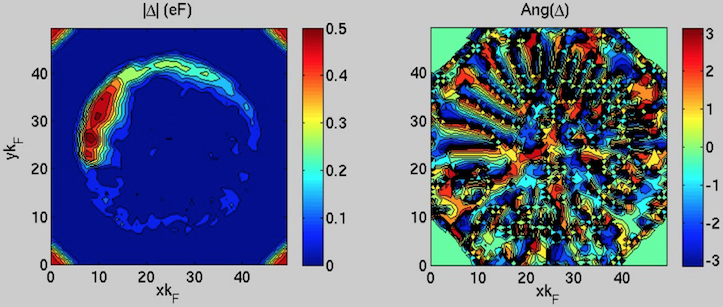}

\caption{\label{fig:can} The first two rows show the magnitude and the phase 
of the pairing gap $\Delta(x,y)$ at various times during stirring.  The position 
of the ``straw'' (which parallel at all times to the axis of the ``can'')  
is visible as small notch in $|\Delta(x,y)|$ at the edge of the ``can'' 
and its linear speed is always smaller than Landau critical velocity. The 
vortex Abrikosov lattice  emerges quite rapidly and the UFG is also partially 
depleted along the rotation axis. If the linear speed of the ``straw'' exceeds Landu critical 
velocity, the order parameter vanishes quite rapidly in time, see the third row.}

\end{figure}
\subsection{Real-time generation and dynamics of quantized vortices.}

The most fascinating applications of the SLDA is to non-equilibrium phenomena. 
In the first simulation of a fermionic superfluid in real time~\cite{Bulgac:2011b} we placed a UFG in a container 
resembling a soda can, homogeneous and with periodic boundary conditions along the longitudinal direction, 
along with other geometries as well.
We inserted a ``straw'' and started stirring the fluid with various constant angular 
velocities and linear velocities  smaller 
and larger than the Landau critical velocity. A UFG, apart from being characterized by 
a very large pairing gap, which in appropriate units is even bigger than in high $T_c$ superconductor~\cite{Magierski:2011}. 
An UFG one has perhaps the largest Landau critical velocity (in appropriate units)  of any 
superfluid~\cite{Combescot:2006,Sensarma:2006,Giorgini:2008}. 
Many of the results obtained are available in the form of videos online, 
for various geometries, various ways to stir the superfluid, and a range of 
stirring velocities ranging from very slow to well above the Landau critical velocity~\cite{Science:2011}. 
If one introduces an object and moves it relatively slowly through the superfluid and 
eventually extract that object slowly also the UFG returns practically to its initial state, as one 
would have naturally expected for an adiabatic evolution. We have also noticed that sometimes 
we can bring the atomic cloud into rotation even if the linear speed of the ``straw'' exceeds Landau 
critical velocity and the cloud remained superfluid. This is possible since UFG is gas, during rotation 
accumulates along the wall, the density and therefore the local Landau critical velocity increases. In 
the same work~\cite{Bulgac:2011b} we have demonstrated in a real-time treatment for the first time 
that in a fermionic superfluid quantized vortices can cross and recombine, exactly as 
Feynman~\cite{Feynman:1955} envisioned and suggested that quantum turbulence emerges, 
see also subsection \ref{sec:qt}.

\subsection{Quantum shock waves.}

\begin{figure}
\includegraphics[width=0.45\textwidth]{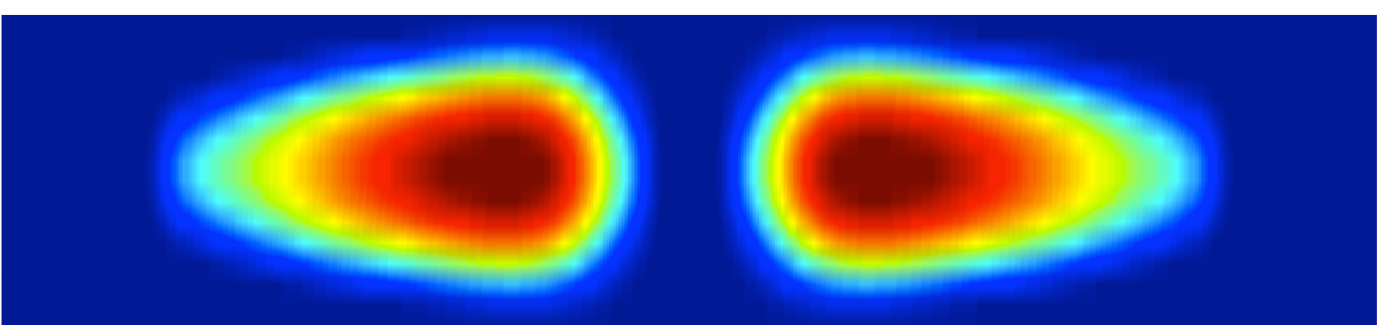}%

\includegraphics[width=0.45\textwidth]{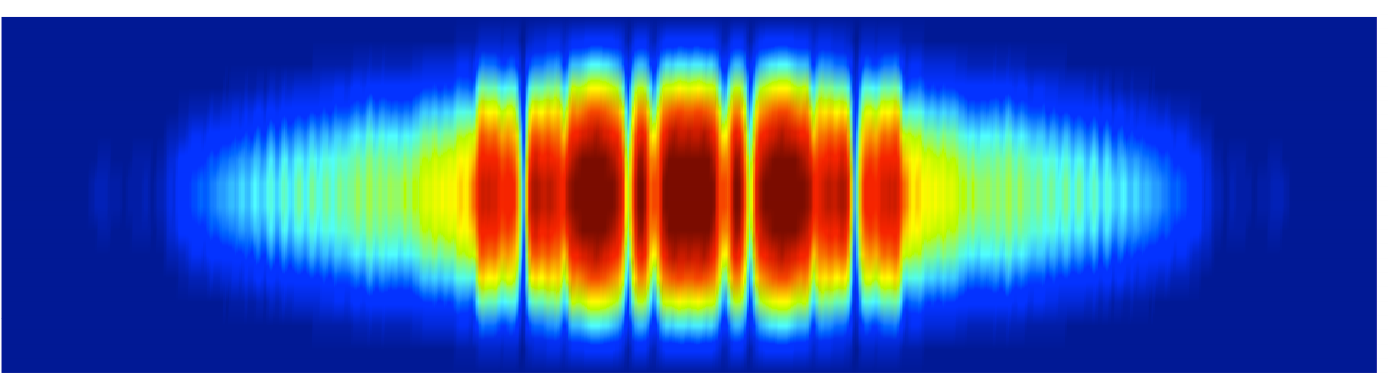}%

\includegraphics[width=0.45\textwidth]{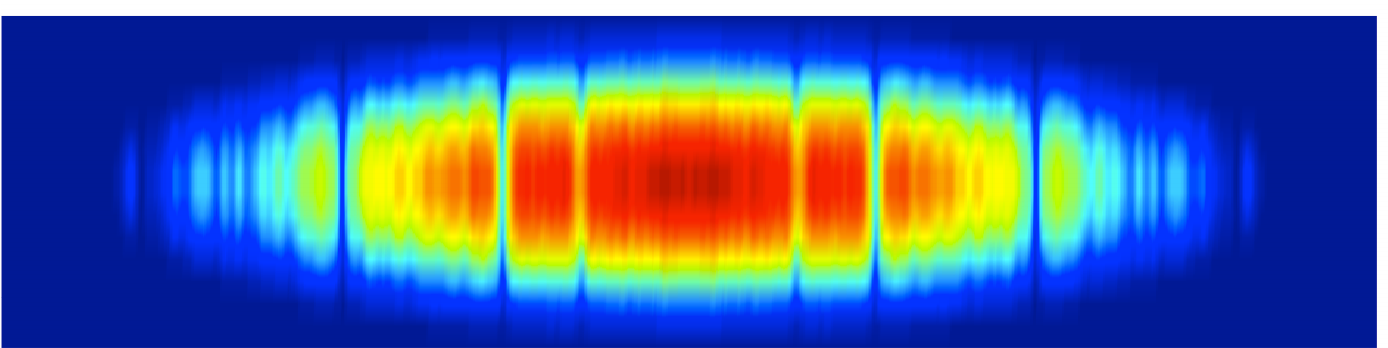}%
\caption{    \label{fig:shock} Three sequential frames of the number density. 
The speed of the front of the shock wave in experiment and simulations agree, though not perfectly. 
The quantum shock wave form after two clouds collide and start expanding, 
but in their wake one can see several density depletions, which are due to 
the excitation of the domain walls~\cite{Bulgac:2011c}.  }
\end{figure}

Thomas and collaborators demonstrated that quantum shock waves can be excited 
in a cold atomic fermionic cloud~\cite{Jpseph:2011}. Thee shock wave front  was directly  
visualized as a clear number density discontinuity. 
In a TDSLDA simulation of this experiment~\cite{Bulgac:2011c} is was shown that the 
character of these shock waves is controlled by the dispersive effects and 
not dissipative effects as was assumed in the analysis of the experiment~\cite{Joseph:2011}. 
At the front the shock wave both the number density and the velocity field are discontinuous, 
when coarse grained appropriately. At the front of a shock wave the matter 
flows in opposite directions.  But in addition with quantum shock waves also domain 
walls are formed, which propagate thought the cloud, see Ref.~\cite{Bulgac:2011c} 
and Fig.~\ref{fig:shock}. A domain wall is the region where 
the phase of the order parameter changes from $-\pi$ to $\pi$ and is topological 
in character, similar to domain walls in magnetism. At a domain wall number density 
has a significant depletion, as in a vortex core.

\subsection{Vortex ring versus heavy soliton.}

\begin{figure}
\includegraphics[width=0.3\textwidth]{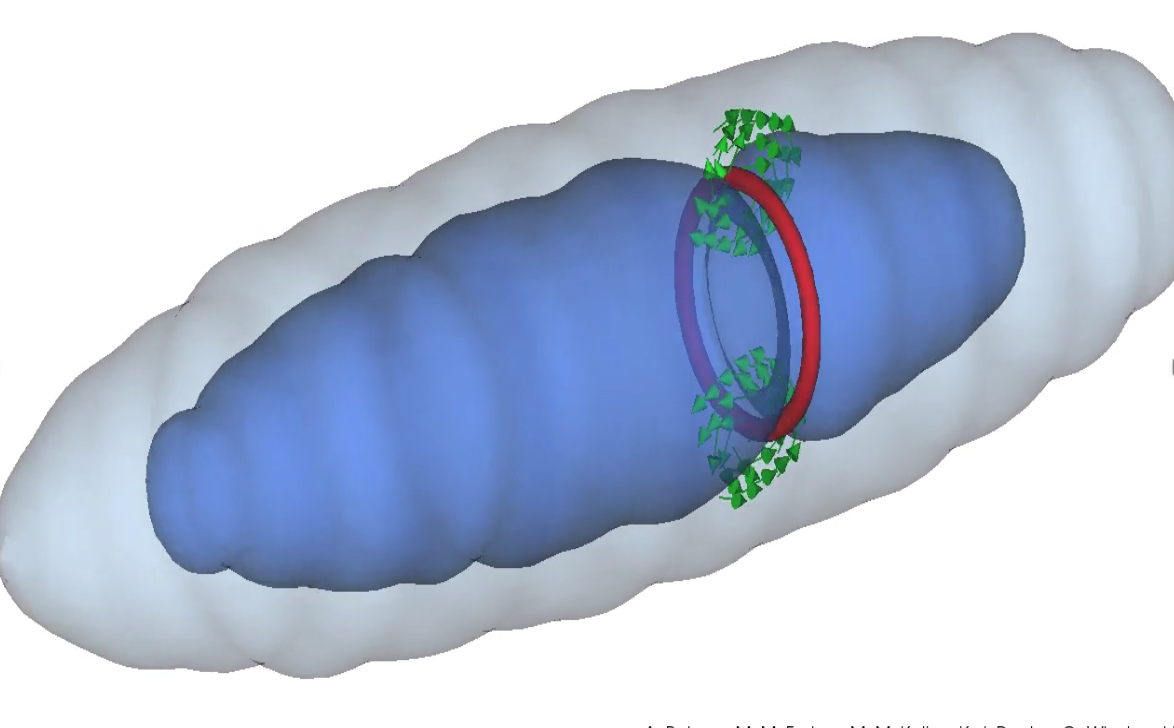}%

\includegraphics[width=0.3\textwidth]{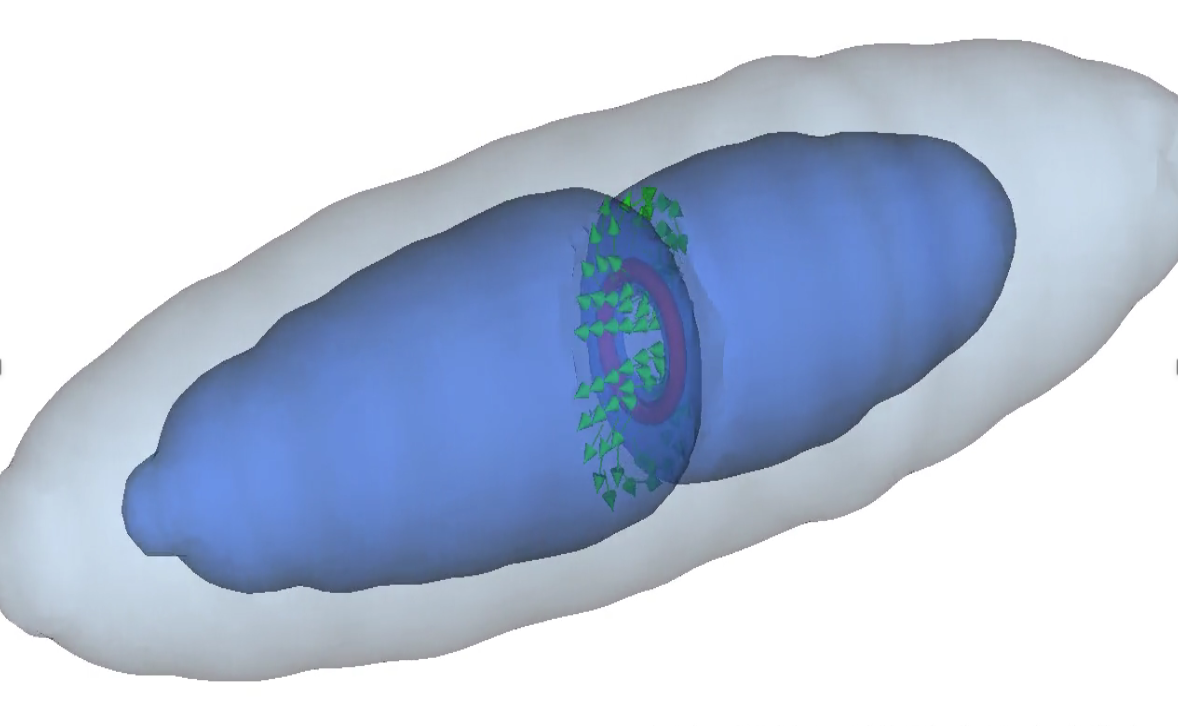}%

\caption{    \label{fig:ring}
After a planar soliton is created by illuminating with a laser half of an elongated fermionic superfluid a 
vortex ring is formed. This vortex ring propagates in one direction with a larger 
radius and in the opposite direction with a smaller radius~\cite{Bulgac:2013d}.}
\end{figure}

In a surprising experiment performed in 2013~\cite{Yefsah:2013} it was announced 
that new kind of  many-body excitation was observed in an 
experiment performed on an UFG. Half of a very elongated superfluid cloud was illuminated 
with a laser, which resulted in  a difference in the phase of the order parameter between 
the two halves. Theoretically is was known for quite some time that in this case such a 
planar soliton propagates with a know speed, and in 3D can be unstable due to the snake-instability.
What the MIT experiment established was this the excitations they created was moving with 
a speed two orders of magnitude slower than the theory predicted and the dubbed this mode a ``heavy soliton.''
In a very careful TDSLDA analysis~\cite{Bulgac:2013d}  we established however, that under the conditions described in 
the original paper~\cite{Yefsah:2013} most likely the authors observed a vortex ring. A planar soliton in an inhomogeneous cloud 
quite rapidly evolves into a vortex ring in simulations and it starts propagating 
back and forth in an almost perfect harmonic motion. However, while moving in one 
direction the vortex ring is large, and while is moving in the opposite direction the vortex ring shrinks, and then the motion 
is repeated, see Fig.\ref{fig:ring}. In subsequent more detailed experiments~\cite{Zwierlein_prl2014,Zwierlein_prl2016} the 
MIT group confirmed the transformation of the planar soliton, into a vortex ring. However, since their trap 
lack azimuthal symmetry, which was not adequately established in the original paper, a vortex ring in an 
axially non-symmetric trap rather quickly touches the walls and turns either into a single or 
two line vortices, dubbed however or solitonic vortices, if vortex line is perpendicular to the propagation direction. 
This is fully in agreement with earlier simulations of an UFG~\cite{Bulgac:2011c}, 
see supplemental material~\cite{Science:2011},  and subsequent 
analysis of the new MIT experiment~\cite{Wlazlowski:2015,Bulgac:2017a}.
 
\subsection{Quantum turbulence.}\label{sec:qt}

\begin{figure}
\includegraphics[width=0.45\textwidth]{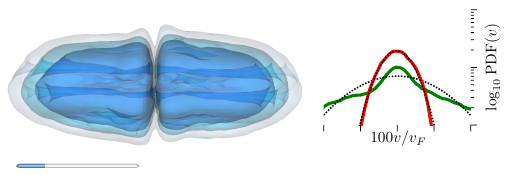}%

\includegraphics[width=0.45\textwidth]{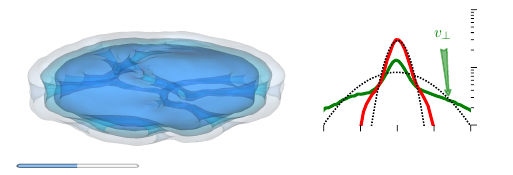}%

\includegraphics[width=0.45\textwidth]{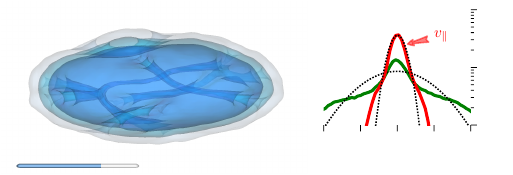}%

\caption{    \label{fig:turb} 
Three consecutive frames illustrating the 
crossing and recombination of quantized vortices in a cold atomic Fermi superfluid are shown in the left column. 
In the right column the longitudinal (red)  and transversal (blue) momentum distributions 
compared with a thermal momentum distribution in logarithmic scale. }
\end{figure}

At the hearty of quantum turbulence lie the formation of a tangle many quantized vortices. 
In experiments with liquid helium such tangles have been created in laboratories for decades~\cite{Vinen:2002}. 
In the case of cold fermionic gases vortices can be easily generated with rotating laser beams~\cite{Zwierlein:2005}
but also by simply illuminating part of a cloud with a laser~\cite{Yefsah:2013}. By combining these 
two methods, which clearly are not the only possibility, one can generate a tangle of quantized vortices
in both unpolarized~\cite{Bulgac:2013d,Bulgac:2017a} and Fig.~\ref{fig:turb}, and in polarized systems~\cite{Wlazlowski:2018}.    
The great advantages of cold atom system over liquid helium are multiple: i) one can control easily many parameters of 
the system, including the interaction strength, and create a wide array of external probes~\cite{Zwerger:2011}; 
ii) one can describe with very good accuracy theoretically both static properties and particularly the non-equilibrium 
dynamics of such systems with very good control of the theoretical ingredients. being able to confront theory and experiment 
in great detail is a great advantage of cold atom systems over studies performed in liquid helium, where 
only phenomenological models exist. 

\begin{figure}[h]%
\includegraphics*[width=0.53\textwidth,height=7cm]{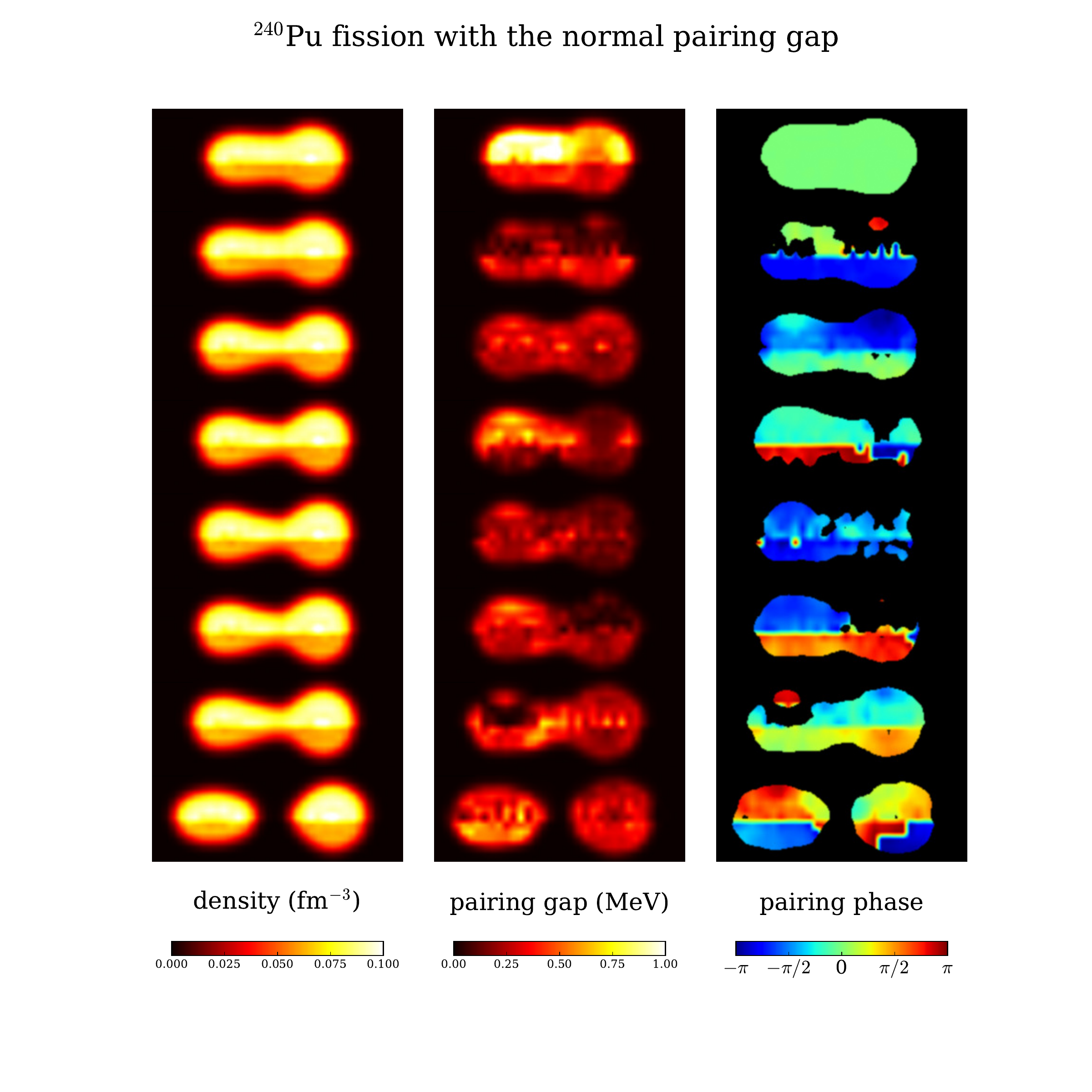}
\caption{\label{fig:fission} The three column show the evolution of the 
neutron/proton number density, magnitude and phase of the pairing respectively 
(upper/lower half of each frame), from the top of the outer fission barrier, past scission, 
until fission fragments are separated.
}
\end{figure}

\subsection{Nuclear fission.} 
Our main goal in developing SLDA was to describe non-equilibrium dynamics of nuclear systems, 
but in order to  verify and validate the theoretical framework and the complex numerical implementation 
we made quite a long detour studying cold atoms systems, for which theoretical tools are both simpler 
and in better control, and also there is a great wealth of experimental data which can be confronted with
theory \emph{predictions} and \emph{postdictions}! One of the oldest problems of strongly interacting  quantum many-body 
systems i s nuclear fission. Unlike superconductivity for example, which required less than five decades 
form the initial  experimental observation~\cite{Onnes:1911} in 1911, 
until a microscopic theory was put forward in 1957~\cite{BCS:1957}, nuclear fission observed in 1939~\cite{Hahn:1939} 
will turn 80 years old in 2019 and likely will still have no adequate microscopic description.  
There are many reasons why this is the case and here are some of them: i) the nuclear interactions 
are extremely complex and they are not yet 
accurately known, unlike Coulomb interaction between electrons and nuclei; ii) nuclei are finite systems and at 
the same time they have too many particles, correlations are very stroong; 
iii) when a heavy nucleus fissions the number of final channels is in 
the hundreds, corresponding to various possible splittings of protons and neutrons between the fission fragments, 
the fission fragments emerge excited with various quantum numbers; iv) apart from fission fragments typically 
quite a number of neutrons are emitted, during fission, immediately after fission, 
and much later, along with gamma rays and beta decays. One cannot declare ``victory'' until theory is able to predict
with reasonable confidence most of these fission fragment properties, which are critical for many applications as well, 
and also for clarifying many fundamental questions, such as the origin of elements in the Universe, and the 
structure and evolution of stars.

Using TDSLDA and an NEDF of reasonable quality in 2016 we were able for the first time for describe 
the evolution of a fissioning nucleus from the outer fission barrier to scission, until fission fragments 
were separated~\cite{Bulgac:2016a} and Fig.\ref{fig:fission} and recently also in Ref.~\cite{Bulgac:2018a}. The fission fragments emerge
  with properties similar to those determined experimentally, while the
  fission dynamics appears to be quite complex, with various shape and
  pairing modes being excited during the evolution. The time
  scales of the evolution are found to be much slower than previously
  expected and the role of the collective inertia in the dynamics is
  found to be negligible.
  Even though in this first study of its kind we did not obtain a
perfect agreement with experiment, our results clearly demonstrate
that rather complex calculations of the real-time fission dynamics
without any restrictions are feasible and further improvements in the
quality of the NEDF, and especially in its dynamics properties, can
lead to a theoretical microscopic framework with great predictive
power. TDSLDA will offer insights into nuclear processes which are
either very difficult or even impossible to obtain in the
laboratory.

\begin{figure}[h]

\includegraphics[width=0.45\textwidth]{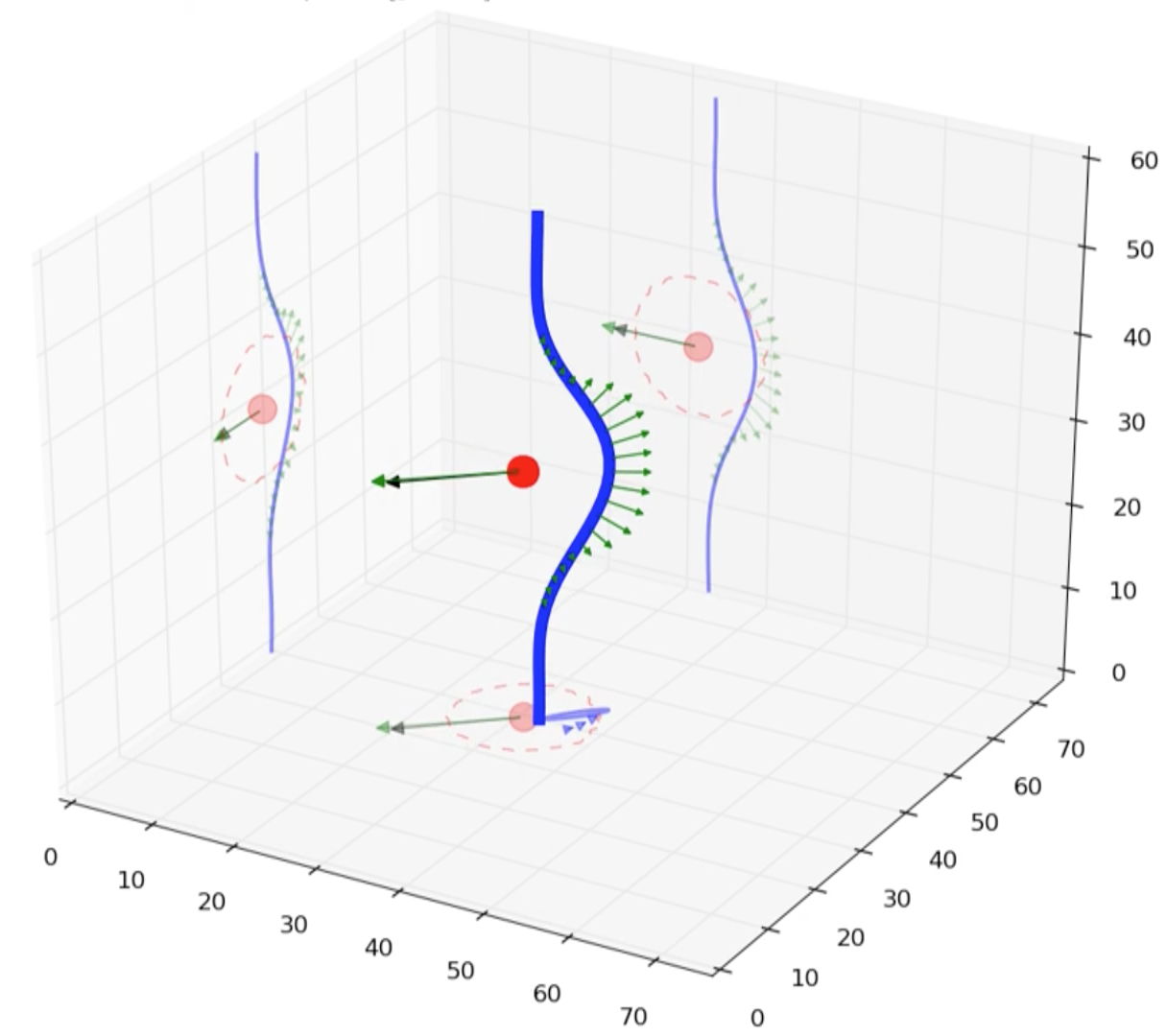}%

\caption{    \label{fig:pinning}
A nucleus moving at constant velocity repels the quantized vortex line. The little green 
arrows display the relative magnitude and direction of the repulsive force between the nucleus 
and the small linear element of the vortex line, while the long green line is the force exerted by the vortex on the nucleus.. }
\end{figure}

\subsection{Vortex pinning in neutron star crust.}

A rather old puzzle in nuclear astrophysics is to explain why neutron stars/pulsars experience glitches. 
Neutron stars are the most compact  in the Universe (not counting black hole, which  are not made of typical matter) 
with a density reaching $\approx 10^{14}g/cm^3$ and they rotate extremely fast, with periods as short as milliseconds. 
As they evolve they radiate energy and their rotation period increases very slowly. However, at practically random 
times they experience sudden accelerations and their periods increase very rapidly, practically instanteneously.
Anderson and Itoh~\cite{Anderson:1975} suggested that since neutron stars are mainly superfluid neutron matter, 
there are many quantized vortices, which likely are pinned in the so called neutron star crust. 
Neutron stars are not composed entirely of neutrons, which are unstable against beta-decay, 
but also of a relatively small fraction of protons and electrons. Electrons are relativistic, very delocalized, 
and likely in a normal phase. But the neutrons and protons lead to the formation of the so 
called pasta phase, a series of nuclei ranging in shape from small spheres 
(meat balls), long sticks (spaghetti) , plates (lasagna), and ``swiss cheese-like'' matter, 
with holes like bubbles and tubes. These ``nuclei'' from a Coulomb lattice in a superfluid neutron superfluid. 
The neutron superfluid under rotation is threaded with quantized vortices, 
which might attach to nuclei or interstitially. During the neutron star rotation these vortices can unpin and find a new position, 
which leads to a change in the star moment of inertia and therefore its period, or glitches, in analogy 
with flux creeping in hard superconducting magnets according to Anderson and Itoh~\cite{Anderson:1975}.

For decades nuclear physicists and nuclear astrophysicists attempted to figure out 
whether a quantized vortex is pinned on a nucleus or whether it is repelled by a 
nucleus in the neutron star crust, and the results have been all over the map, 
from attraction to repulsion, with no clear indication even on the magnitude 
of this interaction. In 2013 we suggest a new approach to this problem, which could 
simply indicate whether the interaction is either repulsive or attractive. A nucleus scattering 
on a quantized vortex would be deflected either to right or to the left depending on the character of the force~\cite{Bulgac:2013a}.
The size and sign of the nucleus vortex interaction arises in any theoretical approach as a difference between 
two very large quantiles, which cannot be computed with enough accuracy.  In 2016 
we have implemented this recipe~\cite{Wlazlowski:2016}, see also Fig. \ref{fig:pinning}  
and we were able to extract   even more finer details of the nucleus-vortex interaction, namely 
how each small linear element of the vortex line interacts with the nucleus. During their interaction the vortex line is 
deformed and stretched. We have established not only that the interaction between a nucleus and a vortex is repulsive, 
for all possible densities on the neutron matter, but we have extracted with quite high accuracy the magnitude of this interaction.

\begin{figure}[h]
\includegraphics[width=0.45\textwidth]{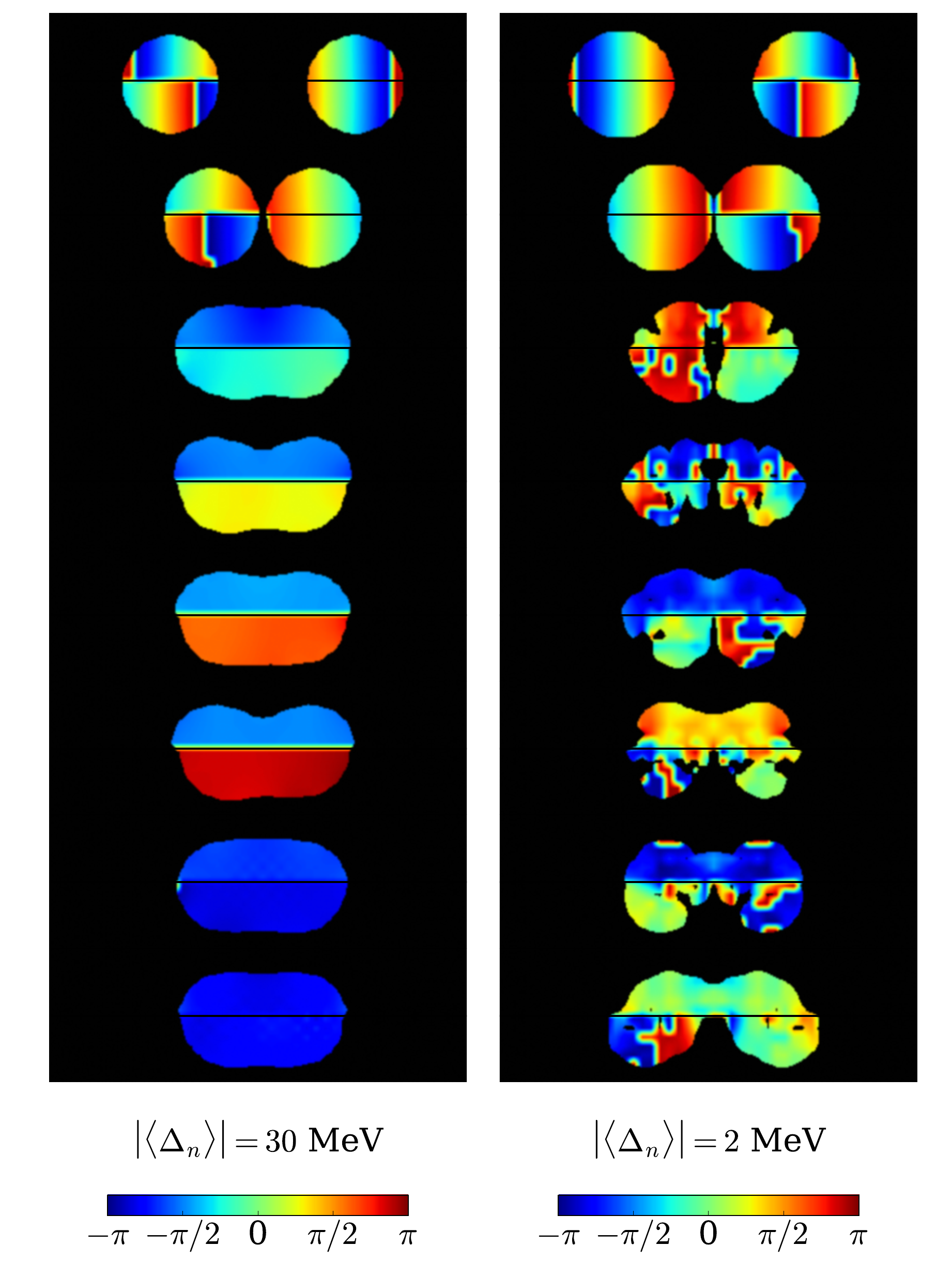}%
\caption{    \label{fig:phase}
The evolution of the phase of the pairing
field (time runs top to bottom) in the head-on collision of
$^{120}$Sn+$^{120}$Sn.  The right and the
left columns correspond to a realistic or artificially increased
pairing field strength respectively. The upper and lower half of each
frame corresponds to an initial phase difference between the two
initial pairing condensates of 0 and $\pi$ respectively.  The phase locking of the pairing
field is clearly manifest after fusion in the left column, but absent
in the right column.  }
\end{figure}

\subsection{Collisions of heavy nuclei and phase locking.} 

I will address only a single aspect of this rather wide field, which 
likely is of interest outside nuclear physics, the phase locking between two colliding 
superfluid droplets. 
Since the gauge symmetry is spontaneously broken in superfluids, it is
reasonable to wonder under what conditions the relative phase of two
superfluids is physically relevant. The Josephson
effect~\cite{Josephson_rmp1964,Josephson_rmp1974}, experiments with
cold Bose or Fermi atoms~\cite{Shin:2004,Yefsah:2013,Zwierlein_prl2014,Zwierlein_prl2016,Bulgac:2014,Wlazlowski:2015}, and the superfluid fragments emerging from
nuclear fission~\cite{Bulgac:2018}, are
just a few examples where that is the case.

Recently Magierski, {\it  et al.}~\cite{Magierski:2017} reported on a surprising and
strong dependence of the properties of the emerging
fragments on the relative phase of the pairing condensates in the
initial colliding nuclei. Related theoretical issues in superfluid 
helium have been discussed in literature by Anderson~\cite{Anderson:1986}.
In Ref.~\cite{Bulgac:2017} we have demonstrated that this is typical behavior of superfluids in the weak coupling limit. 
With increasing interaction strength however, the
initially independent phases of the two order parameters in the
colliding partners quickly become phase locked, as the strong coupling
favors an overall phase rigidity of the entire condensate, and upon
their separation the emerging superfluid fragments become entangled, see Fig.~\ref{fig:phase}.

\subsection{Fluctuations and dissipation.}
 
One of the main difficulties with TDDFT is that in a time-dependent framework is likely to 
always lead to the same final state, even if one were to consider 
a range of different initial conditions\cite{Bulgac:2018a}. As I discussed briefly above, at least in 
the case of nuclear fission and nuclear reactions in general that is definitely not the case.
In nuclear fission one has a wide distribution of fission fragments masses and charges. 
Can one find within TDDFT a sensible solution to this problem? 
 
The prevalent approach to perform real time evolution of many-nucleon systems 
was developed in the 1970's using path-integral techniques; see 
\cite{Negele:1988} for a review. Starting with a system described by a 
many-body Hamiltonian $\hat{H}$, one performs a Hubbard-Stratonovich 
transformation on the many-nucleon evolution operator by introducing auxiliary 
one-body fields $\sigma(t)$, formally,
\begin{eqnarray}
\hat{U}(t_i,t_f) & = & \exp\left [ -\frac{i}{\hbar}\hat{H}(t_{f} - t_{i})\right ] 
\nonumber\\
& = & 
\int \mathcal{D}[\sigma(t)] W[\sigma(t)] 
\exp\left( -\frac{i}{\hbar}\int_{t_{i}}^{t_{f}} \hat{h}[\sigma(t)]\right),
\nonumber
\end{eqnarray}
where $\mathcal{D}[\sigma(t)]$ is an appropriate measure depending on all 
auxiliary fields, $W[\sigma(t)]$ is a Gaussian weight and 
$\hat{h}[\sigma(t]$ is a one-body Hamiltonian built with the auxiliary one-body 
fields $\sigma(t)$. 

Using the stationary phase approximation, one selects a single mean field 
trajectory $\bar{\sigma}(t)$, which one may simulate with the TDDFT trajectory. 
If the initial state is a (generalized) Slater determinant, the final state is 
also a (generalized) Slater determinant under the evolution of this stationary 
phase mean field trajectory. After a trivial change of integration variables 
$\sigma(t) = \bar{\sigma}(t) +\eta(t)$, the true many-body wave function can be 
put into the form
\begin{eqnarray}
&& \Psi(t) = \label{eq:path} \\
&& \int \mathcal{D}[\eta(t)] \bar{W}[\eta(t)]
\exp\left ( -\frac{i}{\hbar}\hat{h}[\bar{\sigma}(t) +\eta(t)]\right)\Psi(0), 
\nonumber
\end{eqnarray}
where $\Psi(0)$ is the initial wave function, and $\eta(t)$ are fluctuations 
around the stationary phase trajectory $\bar{\sigma}(t)$ at time $t$. 
In Eq. \eqref{eq:path} the weight functions are Gaussian-shaped. Thus, the 
true many-nucleon wave function is now a time-dependent linear superposition of 
many time-dependent (generalized) Slater determinants. In this respect the true
many-nucleon wave function has a similar mathematical structure as the wave 
function in the time-dependent generator coordinate method (TDGCM) introduced 
by Wheeler {\it et al.}~\cite{Hill:1953,Griffin:1957}. 
One 
cannot but see the analogy in treating fluctuations around the mean field 
trajectory with the classical Langevin description of nuclear collective motion 
as well~\cite{Frobrich:1998}. 
The representation  \eqref{eq:path} (which is an exact one)  
of the many-body wave function has the 
great advantage that
each trajectory is independent of all the others.
One particular aspect of this general structure of the many-nucleon wave 
function is the nature of the initial wave function. One choice is a single 
(generalized) Slater determinant and another is a superposition of many such 
(generalized) Slater determinants. 

In our fission studies we observed that one-body dissipation (i.e. the 
collisions of fermions with the moving surface of the nucleus) or Landau type 
of dissipation leads to a  very quick dissipation of the f=collective flow energy 
into intrinsic/thermal of the fissioning nucleus~\cite{Bulgac:2018a}.
The intrinsic motion of the 
descending  nucleus from the outer saddle towards the scission configuration is similar to the downward motion of a 
heavy railway car on a very steep hill with its wheels blocked. The wheels dot not rotate but slip and become extremely hot, 
since almost the entire gravitational potential energy of the railway 
car at the top of the hill is converted into heat and very little of it is converted into collective kinetic energy. In this case, 
the railway car velocity is equal to the terminal velocity. An object attains a terminal velocity when the conservative 
force is balanced by the friction force, the acceleration of the system vanishes and the inertia plays no role in its dynamics.
The motion of the railway car is strongly non-adiabatic. Similarly for a nucleus the flow of energy from 
the collective/shape degrees of freedom is controlled by the entropy of the intrinsic degrees of freedom. 
This energy transfer is to some extend stochastic, as is brownian motion, and can be simulated with 
TDDFT evolution equations augmented to incorporate dissipation
 and fluctuations we introduce have the form
\begin{eqnarray} \label{eq:fd}
&&i\hbar \dot{\psi}_k({\bf r},t)= h[n]\psi_k({\bf r},t) + \gamma[n] \dot{n}({\bf r},t)\psi_k ({\bf r},t)\\
&& -\frac{1}{2} \left [ {\bf u}({\bf r},t)\cdot \hat{{\bf p}}+ \hat{{\bf p}} \cdot {\bf u}({\bf r},t) \right ] \psi_k({\bf r},t)
+ u_0 ({\bf r},t) \psi_k ({\bf r},t), \nonumber
\end{eqnarray}
where $\hat{\bm{p}}=-i\hbar\bm{\nabla}$ (not to be confused with
${\bf p}({\bf r},t)$), the index $k$ runs over the neutron and proton
quasi-particle states and where $\psi_k({\bf r},t)$ are 4-component 
quasi-particle wave functions and $h[n]$ is a $4\times4$ partial 
differential operator~\cite{Bulgac:2016,Bulgac:2018}.  The fields ${\bf u}({\bf r},t)$ and $u_0 ({\bf r},t) $ 
generate both rotational and irrotational dynamics and 
the term proportional to $\gamma$ is a quantum frictional term. In the presence of this 
additional term alone $\dot{E}_\text{tot}\le 0$, as in the case of the presence of 
a classical friction term. This is only a phenomenological solution, 
which  likely will lead to reasonable results and might serve as inspiration.

\section{Conclusions.} I presented an extension of the DFT to superfluid systems 
using a local pairing field and a further extension of this framework to 
time-dependent phenomena using the adiabatic approximation. This static 
extension SLDA can be correlated with {\it ab initio} calculations,
it is strongly constrained by a number of theoretical arguments,  and 
at least in the case of cold atoms it appears to have a quite good accuracy. 
The applications to nuclear phenomena is based to a large extent to a phenomenological
energy density functional, which has a sub-percent accuracy for a large number 
of nuclei and their static properties. The time-dependent extension, which in 
addition is required to satisfy  local Galilean covariance, appears to provide a 
correct description of many non-equilibrium processes in nuclei and neutron stars as well. 
 
\begin{acknowledgement}
  I am grateful to my 
  former students and many collaborators, 
  whose names are in the papers we have published together and referenced here,  and to many 
  of my colleagues with whom I have discussed many of the issues covered in this review.
  The results presented here have been obtained with 
   support by US DOE Grant No.~DE-FG02-97ER-41014 and
  in part by NNSA cooperative agreement DE-NA0003841.  
  Calculations have been
  performed at the OLCF Titan and Piz Daint and for generating
  initial configurations for direct input into the TDDFT code at OLCF
  Titan and NERSC Edison. This research used resources of the Oak
  Ridge Leadership Computing Facility, which is a U.S. DOE Office of
  Science User Facility supported under Contract No. DE-
  AC05-00OR22725 and of the National Energy Research Scientific
  computing Center, which is supported by the Office of Science of the
  U.S. Department of Energy under Contract No. DE-AC02-05CH11231.  We
  acknowledge PRACE for awarding us access to resource Piz Daint based
  at the Swiss National Supercomputing Centre (CSCS), decision
  No. 2016153479. This work is supported by "High Performance Computing 
  Infrastructure" in Japan, Project ID: hp180048. Some simulations 
  were carried out on the Tsubame 3.0 supercomputer at Tokyo Institute of Technology.

\end{acknowledgement}

\bibliographystyle{pss}
\bibliography{frascati}

\newpage

\section*{Graphical Table of Contents\\}
In normal fluids turbulence appears only when viscosity is sufficiently strong. 
While viscosity vanishes in superfluids at zero temperature turbulence is possible
if quantum vortices exist. These vortices while moving thorough the fluid will
cross and recombine, in a manner similar to DNA crossing and recombination, 
leading to a chaotic motion dubbed by Feyman as quantum turbulence.

GTOC image: 
\begin{figure}[th]%
\includegraphics[width=4cm,height=4cm]{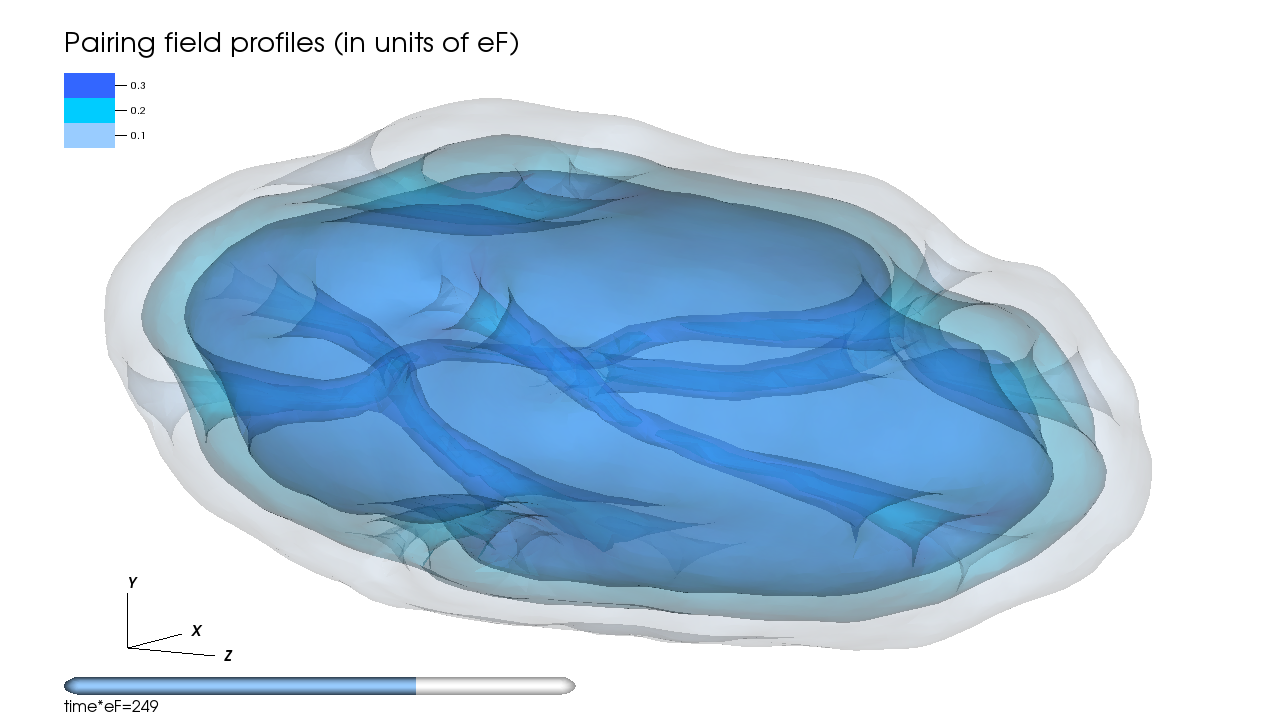}
\caption*{%
A quantized vortex tangle in a cold atomic fermionic superfluid cloud created by bringing 
the cloud into rotation, generating quantized vortices, and by  
imprinting also a domain wall. }
\label{GTOC}
\end{figure}

\end{document}